\documentclass[12pt]{iopart}

\usepackage{graphicx}
\usepackage{array}
\usepackage{fontawesome5}
\usepackage{amssymb}
\usepackage{hyperref}
\usepackage{orcidlink}
\begin{document}

\title[Escape Room with a European Board Game Concept]{Escape Room combined with European Board Game Concepts for self-adjusted Challenge Levels: An educational Eurogame Escape Room in Physics \faPuzzlePiece}

\author{Sascha Albert Bräuninger\,\orcidlink{0009-0008-1809-4101}, Damian Alexander Motz\,\orcidlink{0000-0003-2013-9117}, Matthias Lüpke\,\orcidlink{0000-0001-7855-8301}, Hermann Seifert}
\address{Institute for General Radiology and Medical Physics, University of Veterinary Medicine Hannover Foundation, Bischofsholer Damm 15, 30173 Hanover, Lower Saxony, Germany}
\ead{Sascha.Albert.Braeuninger@tiho-hannover.de}
\vspace{10pt}
\begin{indented}
\item[]\today
\end{indented}

\begin{abstract}
We present an educational escape room in physics (ERP) covering the main disciplines of physics as taught in the education of neighboring sciences, in this case veterinary medicine, extended by a European board game system of victory points (VP). The puzzles in physics are mandatory to master the ERP requiring to be solved in a sequential order. In our study, we show the growth of knowledge by knowledge-transfer cards separated into disciplines of physics as demonstrated by critical exemplary questions of test groups. Tactical secondary puzzles, the so called \textit{Eurogame puzzles} (EPs), are able to be solved in parallel in any arbitrary and independent sequence, allowing a self-adjusted challenge autolevelled by the students by choice. The motivation is improved by a scoring list to solve, at least, a few optional EPs as additional game-like adventure and challenge collecting VP. A second part of the study demonstrates the attraction of the introduced European board game concept as an important tool for gamificiation and open-by-choice EPs not being necessarily related to physics directly. EPs are able to smooth the contrast of playing games and learning, providing an additional bridge to overcome the fear of physics. A strong advantage is the flexibility of the European board concept, choosing from a broad range of optional EPs to fascinate and motivate students taking into account individual preferences improving escape rooms by decision-making skills. The concept is able be applied to every escape room independent of the taught subject or topic.
\end{abstract}

\section{Introduction}
\subsection{General Remarks and Abbreviations}
In everyday life, escape rooms (ERs) and board games are tools of cooperative or competitive ways of analogue entertainment entering the field of education. See Table \ref{table1} for abbreviations as used in this report. The following part introduces briefly the concept of ERs in higher-level education, its advantages and challenges. The second part discusses basics of European board game concepts and the victory point system as used in European board games. This part is important because it is less discussed in the scientific literature, playing a dominant role in the eEERP which combines a sequential ER concept and Eurogame-like ideas to generate an efficient modern concept of gamification.

\begin{table}
\caption{\label{table1}List of abbreviations as used in this publication. To support readability, we have restricted ourselves to use just one or two abbreviations per sentences.}
\centering
\footnotesize
\begin{tabular}{@{}ll}
\br
Abbreviation&Synonym\\
\mr
EP&Eurogame puzzle, Europuzzle\\
ER&Escape room\\
ERP&Escape room of/in physics\\
EER&Eurogame escape room\\
EERP&Eurogame escape room of/in physics\\
PP&Puzzle of physics or physics puzzle\\
VP&Victory point\underline{s} (exceptionally, plural)\\
prefix e&educational (e.g. eER means educational ER)\\
suffix s&plural (e.g. EPs - Eurogame puzzle\underline{s})\\
dc&direct current\\
ac&alternating current\\
\br
\end{tabular}
\end{table}
\normalsize
\subsection{Educational Escape Rooms}
In general, an ER is
\begin{quotation}\textit{a live-action team-based game where players discover 
clues, solve puzzles, and accomplish tasks in one or more rooms in order to accomplish a specific goal (usually 
escaping from the room) in a limited amount of time.} \cite{5e8e8af18402482f89b3c3aedde2703c}
\end{quotation}
Historically, ERs have been built for pure entertainment. Nowadays, the attention has been extended to ERs for the purpose of education as well \cite{5e8e8af18402482f89b3c3aedde2703c,VELDKAMP2020100364,doi:10.1080/20004508.2020.1860284}, called eERs. Game-based eERs have advantages compared to conventional lectures and practical courses promoting
\begin{itemize}
    \item[\faDiceOne] teamwork and collaboration,  social interaction and communication
    \item[\faDiceTwo] higher level enjoyment 
    \item[\faDiceThree] engagement
    \item[\faDiceFour] increased learning gain
    \item[\faDiceFive] critical thinking, problem-solving and creativity
\end{itemize}
being the main advantages \cite{o_szabo_anatomy_2022}. As we will point out in our eEERP, this list is able to be extended by \faDiceSix\, 
 \textit{decision-making and setting priorities} by gauging puzzles.
\subsection{Eurogames}
Eurogames are also called European-style games, European board games or German-style board games \cite{woods_eurogames_2012,reviewBG}. Eurogames are a specific class of board games or table-top games showing characteristics:
\begin{itemize}
    \item Player conflict is indirect,
    \item No player elimination,
    \item Symmetric start conditions,
    \item Score point/victory point system,
    \item Low-level of luck and randomness.
\end{itemize}
The latter points are important, supporting an intellectual multiplayer challenge realised by a contest of collecting score points. Table \ref{eurogames} shows selected Eurogames and details, namely the publisher, game mechanics and the associated number of score points typically received in the game per match. \textit{Settlers of Catan} is a famous example of a Eurogame although the start conditions are slightly asymmetric. One should keep in mind that, the definition of Eurogames is not that rigid and sharp from an academic point of view as expected. Usually, Eurogames are thematic games, mostly not related to real events of history (in contrast to historical board games). Sometimes elements may be mixed, e.g., \textit{Imperial} uses a map and fractions of the First World War. Therefore, Eurogames are not as abstract as chess or checkers. Surprisingly, the aforementioned characteristics (see bullet points above) do not contain a real board as a requirement and the description European-style board game should be understood as a concept. \textit{Carcassonne} is a pure Eurogame although it contains tiles working by placing tiles on the table instead of having a conventional board. In Eurogames, an asymmetry develops mostly over time, e.g. in \textit{Puerto Rico} and \textit{Caylus} by buying buildings. Interestingly, playing the cardgame Skat with score points fulfills a few of the aforementioned points, especially, one can argue the high asymmetry (2 versus 1) is averaged out because a match contains several \textit{hands}. However, the amount of luck is a matter of discussion which depends on the number of played hands and the missing theme (French-suited deck) ends up with the result that Skat is just a Eurogame-like cardgame being extremely on the border to be like. One could try to extend this statement to Rummy or Canasta, but various subgames in Skat (Null, $\clubsuit,\spadesuit, \heartsuit, \diamondsuit$, Grand), are definitively missing deepness and higher randomness in Rummy and Canasta excluding this games. This shows the high complexity of the topic and fluent transition. The most important and clear criterion for exclusion is player elimination which is prohibited in Eurogames. VP of various matches are not necessarily comparable because of the interaction of players and match-dependent dynamics. Additionally, start conditions may be match-dependent. \footnote{Issues of Eurogames are rarely discussed in the scientific literature. The main part of this subsection roots in decades of experience playing Eurogames and visits to the \textit{Internationale Spieletage SPIEL}/Essen Game Fair (world's greatest board game fair) of the main author.}
\begin{table}
\caption{\label{eurogames}Selected Eurogames using a victory point system to identify the best player of the match. Typical VP are mentioned, being strongly dependent on the number of players and dynamics of the match.}
\centering
\footnotesize
\begin{tabular}{@{}llll}
\br
Eurogame&Publisher&Mechanics&typical VP\\
\mr
\textit{Settlers of Catan}&\textit{Kosmos}&Dice rolling&max. 10\\
\textit{Carcassonne}&\textit{Hans im Glück}&Area majority, tile placing&70-125\\
\textit{Puerto Rico}&\textit{Alea}&Action drafting&25-65\\
\textit{Caylus}&\textit{Ystari Games}&Worker placement&80-125\\
\textit{Imperial}&\textit{PD Games}&Investment, area majority&75-180\\
\br
\end{tabular}
\end{table}
\normalsize
\section{Eurogame Escape Room in Physics}
\subsection{Goal, Room Concept, Capacity and Restrictions}
Coming back to the above definition of an ER, the specific goal of our ERP was not to escape from the room. We used an ER concept of two rooms. The ER started in the first room (so-called mirror room, starting room) and the intermediate goal was to enter the second locked room which was named Schrödinger's working room, see S. A. Br\"auninger \textit{et al.} \cite{tiho_mods_00010090,tiho_mods_00010496}. The entrance required an opening of eight boxes by solving the first eight PPs. Therefore, our ER was an inverted ER taking into account fire protection issues as stipulated by the university. The main goal was to finish all PPs. The secondary (not mandatory) goal was to collect VP by solving the puzzles using a victory point system as in Eurogames. The ER was team-based solved by 2-4 students. However, ideally the aimed occupancy was 3 students. The processing time was 2.5 hours. The students are hereinafter called players as well. The instructor is hereinafter called e-game master (educational-game master). The e-game master was available by phone combined with diluted visits to check the work flow. The starting room was not locked to allow the players to use the restroom and mobile phones were allowed to be taken into the eERP. However, digital detoxification (digidetox) was strongly urged. The ERP could be freely chosen as a compulsory elective subject. Approximately 85\% of the students were female (studying veterinary medicine at University at the Veterinary Medicine Hannover Foundation, year 2023). The eERP is a spectral one, which means it covers a variety of main disciplines, in this case in physics (mechanics, optics, electromagnetism: dc/ac, atomic physics, solid states, thermodynamics, X-ray).

\subsection{Rules, Educational Descriptions and Documents}
The ER containes four kinds of documents to support the understanding, providing a transfer of knowledge of physics, atmosphere and instructions to master the puzzles.
\begin{itemize}
    \item \textbf{Story}: The story document contains a story-like introduction to explore the loss of Schrödinger's cat. This document is a cosmetic one for thematic embellishment. It provides an atmosphere without transferring important knowledge to the reader to master the ER. 
    \item \textbf{Rulebook}: The rulebook explains the eurogame system of VP containing a list of the values of points for PPs and EPs. This list can be found in the \textit{Appendix} in table \ref{jlab4} and will be discussed in the following subsection as well. The rulebook discusses the general principle of the ER (opening boxes and treasures by codes, solving the PPs in a sequential order).
     \item \textbf{Knowledge transfers}: Every PP contains a document of 1-2 pages (DIN A4) presenting important knowledge of physics strongly associated with the puzzle. Every asked question in the questionnaire targeted in our presented study could be answered successfully by the transfer documents. The knowledge cards are unassigned, containing one digit for the next code of a box/treasure. Therefore, the players had to combine the PPs and the knowledge transfer documents.
      \item \textbf{Puzzle instructions}: Short supplements of 1 page (DIN A5) contained tips, safety remarks and experimental issues. Puzzle instructions were only added to PPs.
\end{itemize}
The visit to the eEERP started with an audio explanation of a few minutes summarising the game rules. This highlighted important key rules.

\subsection{Victory Point Concept}
The accumulation of VP is one motor of motivation. The effective 'match' of the game is counted over one semester to compare the collected VP of groups of players. This is called a \textit{run}. The students visited the ER collected points in accordance with the Eurogame concept to compare their VP per run. Every group which successfully completed the ER accumulated points for two kinds of puzzles: PPs and EPs. Both kinds of puzzles are explained in the following subsections.\\Coming back to the above definition of an ER, the primary \textit{specific goal} of our eEER was to complete all PPs successfully to master the ER. The secondary goal was to collect VP by solving all PPs and EPs. Table \ref{jlab4} (see \textit{Appendix}) shows the victory points $P_i$ of PPs and points $p_i$ of EPs as used in our ERP in 2023/2024. The points for PPs have two purposes. First of all, without solving any EP the players would end up with 0 VP (after mastering), this being a frustration-like number of points. To avoid this, PPs give points, therefore providing improved motivation. The second reason is to offer support if the students have problems solving individual puzzles in physics. Here, we halved the points to sell ideas. Values of points scaled with the difficulty of the puzzle. Therefore, the maximum of VP was
\begin{equation}
p=\sum_{\mathrm{PPs}}P_i+\sum_{\mathrm{EPs}}p_i=\sum_{i=1}^{10}P_i+\sum_{i=1}^{16}p_i=40+126=166.
\end{equation}
Realistically, it was impossible to solve all puzzles because the EPs were quantitatively overestimated, offering a broad proposal to motivate the players by selecting their individual preferences according to skills, strengths and passion. Thus, the players should focus on the mandatory PPs, solving the optional EPs diluted, i.e., only a few in parallel. 
\subsection{Puzzles of Physics}
\begin{figure}
    \centering
    \includegraphics[width=\textwidth]{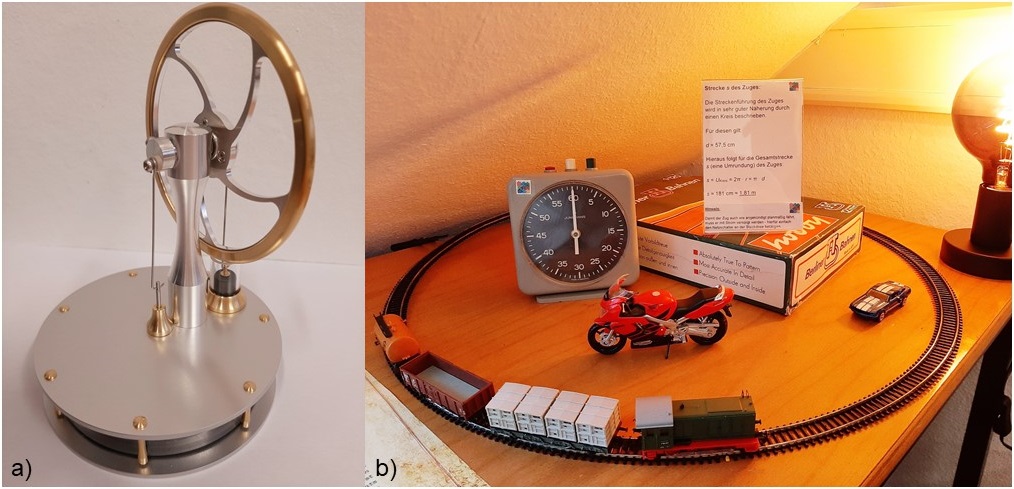}
    \caption{a) Miniature Stirling engine which runs on several heat sources, e.g., cup of hot water, b) train experiment to calculate the velocity of a moving model train on railroads by stopping a grey clock.}
    \label{fig:PP}
\end{figure}
\begin{figure}
    \centering
    \includegraphics[width=\textwidth]{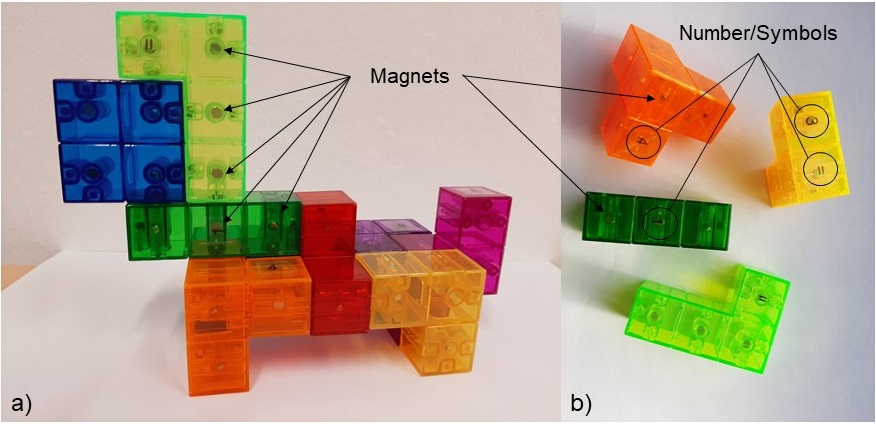}
    \caption{a) dog which consists of tetris-like blocks containing magnets inside to stabilise the construction of this puzzle b) separated blocks emphasizing the added numbers and symbols to achieve equations after puzzling the dog.}
    \label{fig:magnetdog}
\end{figure}
The puzzle structure is \textit{sequential} according to A. Veldkamp \textit{et al.} \cite{VELDKAMP2020100364}. PPs are mandatory to finish the ERP successfully. Table \ref{jlab1} shows the list of PPs and transferred knowledge. In total, 10 PPs were developed. Exemplarily, there are three PPs presented in Figure \ref{fig:PP} and \ref{fig:magnetdog} which are discussed in the following:
\begin{itemize}
    \item[\faPuzzlePiece] \textbf{Stirling engine}: Stirling machine which must be put on a cup of hot water; see Figure \ref{fig:PP} a). Parts of the equipment are hidden. The machine is a final puzzle without generating a new code. The photo shows the \textit{Kontax KS90 Silver Stirling Engine} from the United Kingdom (high quality product). The main knowledge of transfer is that the efficiency $\eta$ of a heat engine is 
\begin{equation}
\eta=\left|\frac{\Delta W}{\Delta Q}\right|=1-\frac{T_2}{T_1}<1
\end{equation}  
where the second equality concerns a Carnot engine. The latter one is mentioned for didactic simplicity being the famous example of literature. Heat $\Delta Q$ can not be transformed completely in mechanical work $\Delta W$.
    \item[\faPuzzlePiece] \textbf{Train experiment}: A miniaturised train was used as puzzle in kinematics, see Figure \ref{fig:PP} b). It drives in a circle and the task was to calculate the velocity and a transformation of units to get the next code for the final treasure. The experiment contains a miniaturized train with rails (scale 1:120) produced by \textit{Berliner
Bahnen TT Hobby} with a trail of 12\,mm. For details, see \textit{Escape room in physics: Train experiment} \cite{tiho_mods_00010208} which contains a description to extend the puzzle by an error calculation. The result of the PP is 
\begin{equation}
  v=\frac{2\pi r}{t}\approx \frac{180.6\,\mathrm{cm}}{16\,\mathrm{s}}\approx11.3\,\mathrm{cm/s}.
\end{equation}
where $r$ is the radius of the circle of around $r=d/2$ with $d\approx57.5$\,cm and $t\approx16$\,s the time per circle of the train. The first two digits "1" and "1" are the first two digits for the code of the next box. The third digit is given by the correct dedicated knowledge transfer card. The players must match all knowledge cards to the corresponding PPs because the knowledge cards are matched. An important issue is to identify a circular motion as accelerated movement in which the direction changes (change of direction of $\vec{v}$). The absolute value is constant: $|\vec{v}|=v$.
    \item[\faPuzzlePiece] \textbf{Magnetic dog puzzle}: Magnetic tetris-like stones/blocks in 3D are used to visualise ferromagnetic forces; \textit{CUBIMAG Pro} produced by \textit{HCM Kinzel} from Germany, see Figure \ref{fig:magnetdog} a). This puzzle has a high gamification and difficulty level. The blocks are stabilised by small magnets (grey pieces) inside the blocks. We added numbers and mathematical symbols ("+","-","="), see Figure \ref{fig:magnetdog} b). After assembling, it yields two digits (two equations). The individual blocks can be turned because of the symmetry recalling symmetry operations in molecules and solid-state physics. This must be considered. Additionally, small experiments are supplemented, a demonstrative add-on without any VP, see \textit{Appendix B}. In our case, the quantity of interest is the Curie temperature $T_C$ of the material and the magnetic susceptibility is
\begin{equation}
  \chi=\frac{C}{T-T_C}
\end{equation}
in which the magnet shows spontaneous magnetization below $T_C$ and room temperature should be understood as $\approx293$\,K\,$=T<T_C$ to adapt the Kelvin-scale as well. $C$ is a material-dependent magnetic constant.
\end{itemize}
\begin{table}
\caption{\label{jlab1}Overview of the ten PPs. Additionally, corresponding topics in the field of physics are presented and important laws or quantitative relations are transferred to the reader by knowledge cards.}
\footnotesize
\begin{tabular}{@{}lll}
\br
Puzzle/Quest in Physics&Topic&Knowledge Transfer\\
\mr
Nobel prize (start-up)&History of physics&Important physicists, Nobel prize modalities\\
Apple puzzle&Mechanics and forces&Newton's law of motion\\
Fluorescence&Atomic physics&Excitation of atoms by UV-light\\
Electric conductivity&Electricity (dc)&Ohm's law\\
Total internal reflection&Geometric optics&Reflection and refraction\\
50 Hertz sound experiment&Electricity (ac)& Frequency $\nu=\frac{1}{T}$ as inverted period time $T$\\
X-ray&Radio diagnostics&Absorption coefficients\\
Dog puzzle&Solid state physics&Force of ferromagnetism, symmetry\\
Train experiment \cite{tiho_mods_00010208}&Kinematics&Mechanical circular motion\\
Cup Stirling machine&Thermodynamics&Heat engine, inverted fridge\\
\br
\end{tabular}
\end{table}
\normalsize
\subsection{Eurogame puzzles}
\begin{figure}
    \centering
    \includegraphics[width=\textwidth]{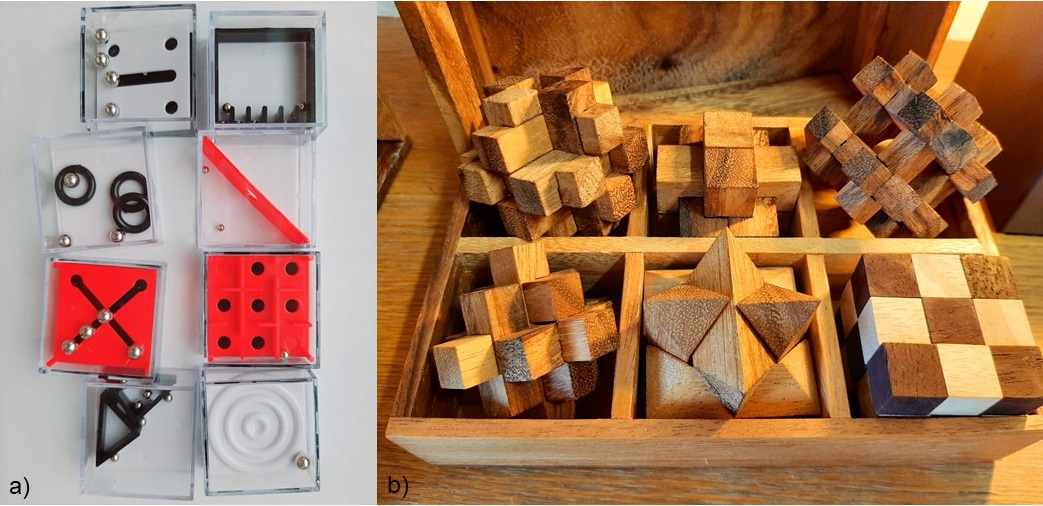}
    \caption{a) Twenty-four dexterity cubes distributed in various boxes in the mirror room matching the modern environment (eight are shown) b) six wood puzzles located in the rustic room called Schrödinger's workroom.}
    \label{fig:EPs}
\end{figure}

EPs are optional and arbitrary puzzles. It is not mandatory to solve EPs, but rather they are an opportunity to collect additional VP. The puzzle structure is \textit{open} according to A. Veldkamp \textit{et al.} \cite{VELDKAMP2020100364}. EPs are independent of each other and independent of the PPs. A few EPs are located in the second room, therefore these puzzles are available later on. An advantage of EPs is the possibility to include commercial and arbitrary puzzles in every ER, in the present case in an eERP, thus offering an optional supplement. EPs do not necessarily have anything to do with the topic of education, in this case physics. To avoid confusion, a EP is not a Eurogame. EPs are arbitrary commercial or self-made puzzles that earn points after solving, comparable to Eurogame elements, e.g. by building a building (\textit{Caylus, Carcassonne}) or by shipping resources (\textit{Puerto Rico}).
Once again, EPs are not mandatory and fully optional puzzles. Exceptionally, the team of players is motivated to earn more VP, thus accepting the challenge. A mark/grade was not given. It was only possible to fail or pass the ERP 'examination'. Table \ref{jlab2} shows the list of the 16 EPs in the eEERP. Exemplarily, three EPs are presented:
\begin{itemize}
    \item[\faPuzzlePiece] \textbf{24 dexterity cubes}: One group of EPs are 24 cubes requiring mechanical dexterity; these were bought from the company \textit{Furado} from China, see Figure \ref{fig:EPs} a). The aim is self-explained by logic and the first part of the challenge to find it out. The second part is mechanical. For instance, it contains four holes inside and this recess should be filled with four metal balls. Each of the 24 EPs has individual challenges promoting surgical dexterity which is needed in veterinary medicine.
    \item[\faPuzzlePiece] \textbf{6 Wood puzzles}: Figure \ref{fig:EPs} b) shows the presented six wood puzzles from \textit{Casa Vivente}. Two sets of the puzzle were bought. One set is already assembled and the second one should be assembled. The 6 EPs require a combination of dexterity and spatial sense/imagination.
    \item[\faPuzzlePiece] \textbf{Magnetic snake puzzle}: The players may build a snake after finishing the dog puzzle as shown in Figure \ref{fig:magnetdog}, using this equipment twice.
\end{itemize}
\begin{table}
\caption{\label{jlab2}EPs supplemented by a short description.}

\footnotesize
\centering
\begin{tabular}{@{}lll}
\br
Number&EBG puzzle&Description\\
\mr
1&Backgammon&UV-letters on the board yielding a name of a physicist\\
2&Biochemistry&Order chemical reaction fragments based on tyrosine and Iodine\\
3&Domino&UV-letters on domino stones yielding a name of a physicist\\
4&Einstein lock&Open the complicated lock puzzle\\
5&Bottle safe&Open the safe made of wood and cord\\
6&24 dexterity cubes&For instance, put metal balls in a hole\\
7&6 Wood puzzle&Build up cubes consisting of small pieces of wood\\
8&Card game&Order cards to yield the name of a physicist\\
9&10 knots&Knot knots following instructions\\
10&Magnet blocks&Build an animal (here a snake) with building blocks\\
11&Periodic table&Abbreviations of elements yielding a name of a physicist\\
12&Wood bottle&Build a bottle made of wooden puzzle pieces\\
13&Chess&Best moves in Scandinavian defence yields a name by letters\\
14&3x3 square&Assemble squares to a large 3x3 square\\
15&5 Text challenges&Logical exercises in envelopes as written text\\
16&Hidden card&A card is very well hidden in a drawer\\
\br
\end{tabular}
\end{table}
\section{Details of the study}
\subsection{Group of interest}
\begin{figure}
    \centering
    \includegraphics[width=0.7\textwidth]{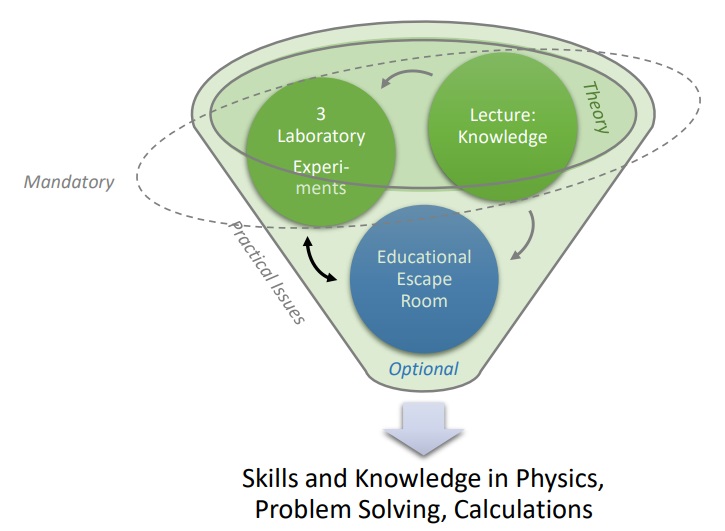}
    \caption{Education in physics at the University of Veterinary Medicine Hannover, Foundation includes visiting lectures, participating in the laboratory course supplemented by the optional eEERP. Arrows visualise dominant transfer of knowledge and skills.}
    \label{fig:trichter}
\end{figure}

The students being educated in physics can be divided into three groups, as follows:
\begin{itemize}
    \item Group A: Students who require a high-end education in physics (main study being physics, biophysics, quantum engineering, ...)
    \item Group B: Students who require an intermediate level of education in physics (main study being in chemistry, machine engineering, mechanical engineering, ...)
    \item Group C: Students who should have a basic knowledge of physics (main study being in medicine, veterinary medicine, biology, ...)
\end{itemize}
The education in physics of these different groups is characterised by specific dedicated and methodological challenges. Empirically, the levels of motivation and interest are low in group C. The target group of our study was group C and the eER was visited by students of veterinary medicine in their first semester aged between 18-22 years. The students had already attended 2/3 of the lectures in physics covering the main disciplines of physics. The majority of the students passed their experimental examination in three laboratory courses: current of liquids, light microscopy/polarimetry and electricity (dc). By choosing the eEERP, the education is organized in three parts, see Figure \ref{fig:trichter}.
\subsection{Educational objectives}
Group C should be taught a basic knowledge of physics and the goals of the eERs are
\begin{itemize}
    \item Stabilisation/Repetition and application of knowledge in physics as learned in the lecture 
    \item Slight extension of the knowledge by exemplification, demonstration and individual cases associated with veterinary medicine
    \item Training of experimental skills and \textit{hands-on-mentality}
\end{itemize}
The latter point is important because it is a matter of interest in veterinary medicine when examining patients. Especially, the corona pandemic led to a strong request for the training of practical skills after the intensive online-teaching. It should be mentioned that because of the broad range of disciplines in physics, a processing time of 2.5 hours and the required intensity of our eEERP should not be used for Group A, at least taking into account the here presented knowledge-transfer documents.
\subsection{Questions and procedure}
The study contains two questionnaires. The first one contains 12 questions on physics from various disciplines which can be found in the \textit{Supplementary}. This questionnaire was answered right at the beginning of the ERP by all players individually, being separated from each other to avoid interaction (called pretest). After the successful completion of the eEERP, they received a blank questionnaire with the identical questions again (called posttest). One example of a question in physics was: 
\begin{quotation}
\textit{What is the single-phase alternating current frequency in the USA?}
\begin{itemize}
    \item[$\square$] 50 Hz
     \item[$\square$] 55 Hz
      \item[$\square$] 60 Hz
       \item[$\square$] I do not know
\end{itemize}
\end{quotation}
We added the answer '$\square$ I do not know' to avoid random choices. Honesty was asked for when filling out the questionnaire. The questionnaires were anonymous. The posttest was supplemented by an additional questionnaire asking questions concerning the VP system, Eurogame and room concept called EER-test. This was the second part of the study. All tests were compiled in such a way to ensure that all answers were given by the same player.
\section{Results}
The evaluation contains two parts. The first part is summary of a set of various straight-forward analyses of unmodified raw data according to statistical methods for statistical significance. The second part presents a demonstrative and simple understandable modified analysis using a strategy to suppress large errors assuming honesty.  
\subsection{Raw data analysis: Test of Significance}
Pretests and posttests were statistically evaluated without further normalisation or modification and, thus, sorted into \textit{raw data}. In the following, this evaluation strategy is be presented as an overview of the main results (for details, see \textit{Supplementary}). The counting of the number of correct, wrong as well as neutral answers formed the basis for the evaluation. Four complementary statistical analysis approaches, in detail 
\begin{itemize}
    \item \textbf{Approach 1} was based on the total number of students $N_1 = 25$,
    \item \textbf{Approach 2} was based on the total number of questions $N_2 = 12$,
    \item \textbf{Approach 3} was based on the group-separated total number of questions $N_3 = 12$,
    \item \textbf{Approach 4} was based on the question-separated total number of student groups $N_4 = 8$,
\end{itemize}
 were performed with the focus on a statistical comparison of the pre- and posttest averaged number of correct answers. To achieve statistically confirmed conclusions, statistical tests in the form of 
 \begin{itemize}
     \item David tests for normal (Gaussian) distribution \cite{gottwald_statistik_2004},
     \item F-tests \cite{gottwald_statistik_2004,funk1992donnevert},
     \item standard two-tailed t-tests (equal variances) \cite{842121},
     \item Grubbs tests for outliers \cite{gottwald_statistik_2004,funk1992donnevert}
\end{itemize}
 were applied to the data. Overall, the evaluations revealed a statistically (highly) significant increase in the averaged numbers of correct answers in the posttest and, therefore, after the students had completed the ERP and proved the success in their physics knowledge transfer by the developed eEERP. Furthermore, the examination also demonstrated that most of the students renounced guessing correct answers in the questions.
 \begin{figure}
    \centering
    \includegraphics[width=\textwidth]{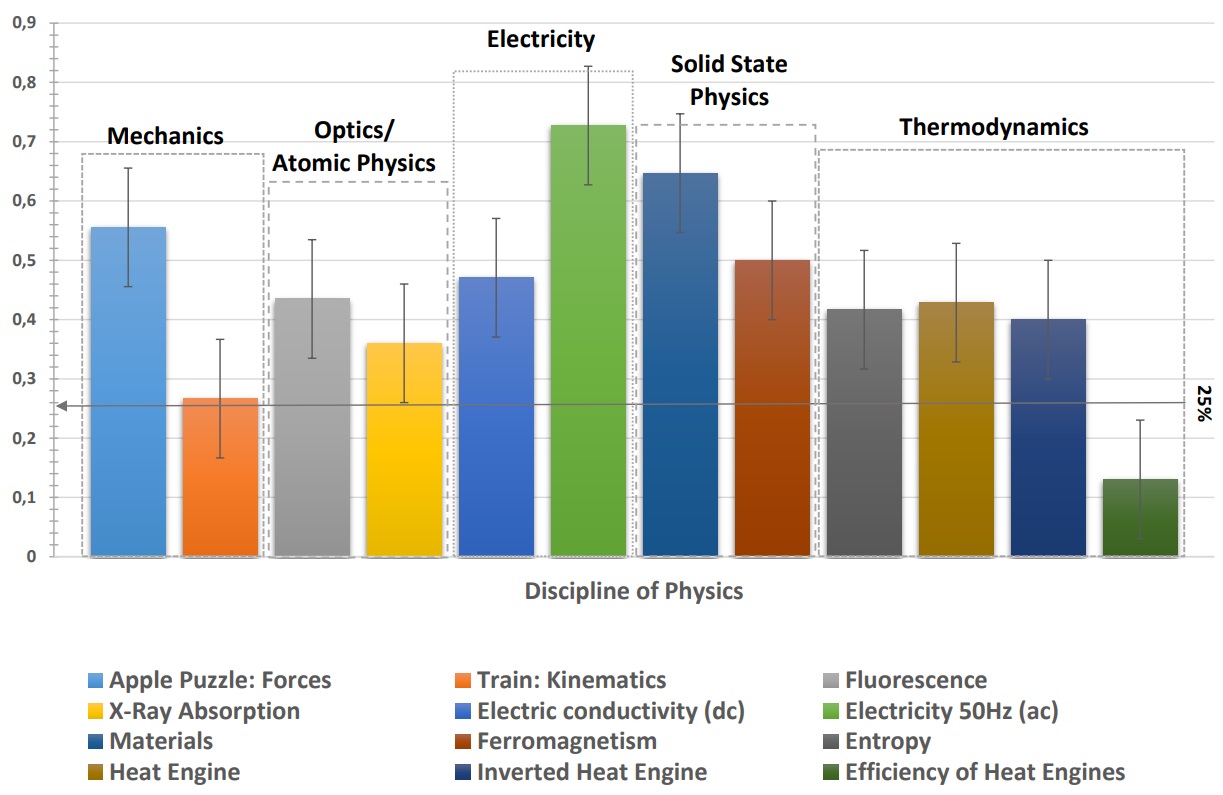}
    \caption{Overview of normalised increase in quantified knowledge of players with bemusement before (only players with incorrect answer or 'I do not know' in the pretest were considered because of measurability). Left scale: 0 means no player has learned anything, 1 means every player has given the correct answer after the eEERP.}
    \label{fig:questions}
\end{figure}
\subsection{Increase in Content Knowledge}
To visualise the increase in knowledge, a sheet of questions is generated divided into disciplines of physics as shown in Figure \ref{fig:questions}. Some players knew the right answer in the pretest, therefore, these answers are not considered in the statistic and pretests and posttests were excluded concerning the corresponding questions/answers. For instance, three players knew (pretest) that the ac-frequency is 60 Hz (USA) and were excluded. Most of the test groups decided '$\square$ I do not know' and after the eEERP a considerable 16 players (72\% or 0.72) chose 60 Hz, these students not knowing the answer before. The possible error has two extremes:
random choice of the answer yields an error of 25\%, meaning just 1/4 of the students chose the correct answer. This assumes random arbitrary decision and is an effective scattered background (see grey 25\%-line). The second extreme is that the players decided honestly, choosing '$\square$ I do not know' if they had truly no answer. In this case, the effective error is zero and no background exists caused by randomness. The truth is something in between with a tendency to the second extreme. Therefore, the presented error was chosen to be 10\% which still overestimates the reality (request for honesty). One can clearly adumbrate that the 'balance point' averaged over all bars is located above the 25\%-line threshold of random choices in the posttest. This is consistent with the above analysis of significance. The mean values are $\approx0.44\pm 0.15$ and $0.47\pm 0.13$ without considering the last question (efficiency of heat engines); for reasons, see \textit{Discussion}.
\subsection{eEER, Room Concept and General Questions}
\begin{figure}
    \centering
    \includegraphics[width=\textwidth]{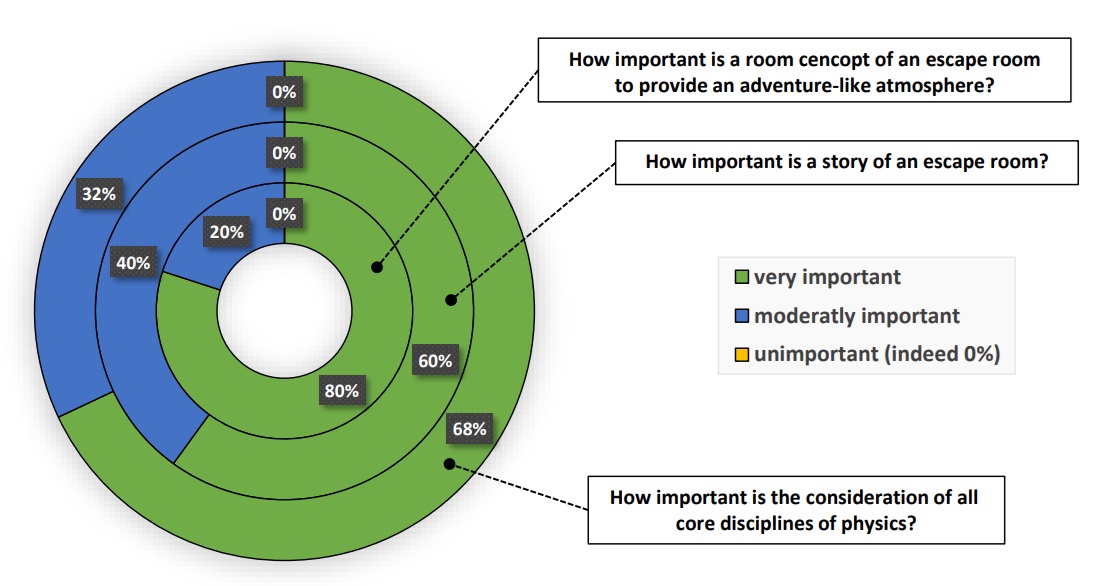}
    \caption{Circle diagram of general questions concerning an ERP. Interestingly, the choice 'unimportant' was not elected.}
    \label{fig:gen}
\end{figure}

\begin{figure}
    \centering
    \includegraphics[width=\textwidth]{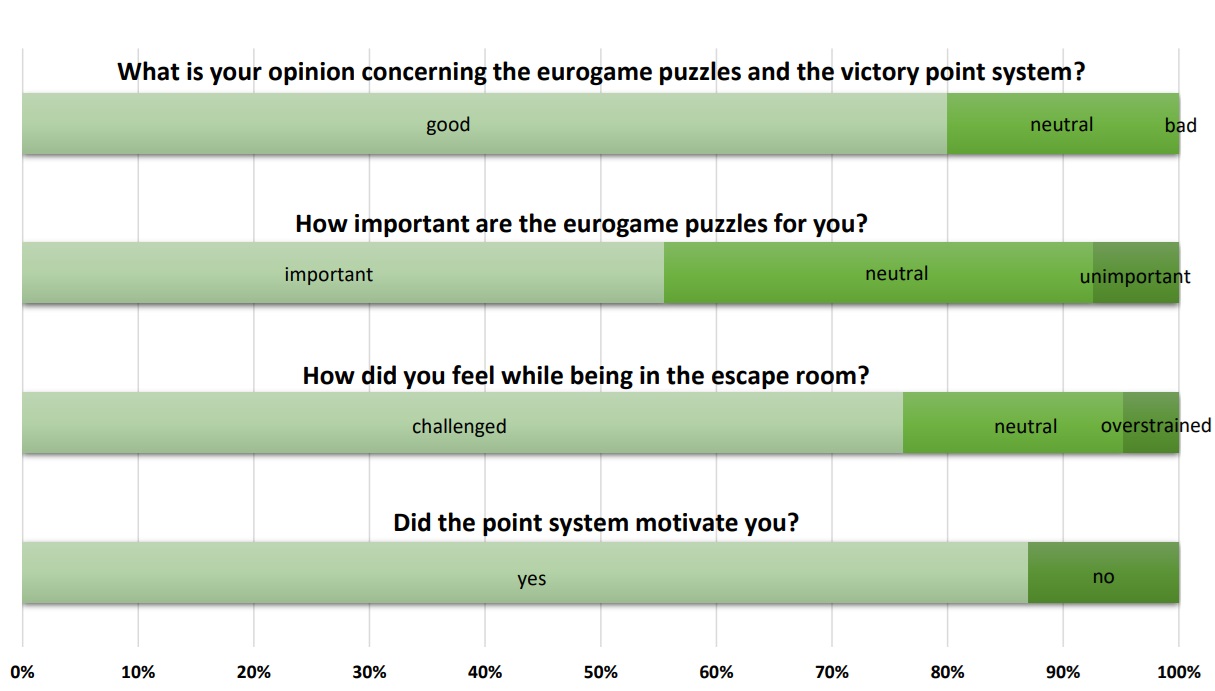}
    \caption{Percentage-bar diagrams of eEER questions.}
    \label{fig:euroq}
\end{figure}

General issues of the eER, the level of acceptance and quality of the Eurogame concept were investigated by the EER-test. Figure \ref{fig:gen} shows three of the rated questions. A total of 80\% rated the existence of a room concept (mirror room, rustic Schrödinger's workroom) to be very important to provide an adventure-like atmosphere. Interestingly, no player rated an element of interest to be unimportant. This is surprising because the story of an escape room is cosmetic and wastes time to finish the ERP. However, 40\% of the respondents evaluated it to be just moderately important. A considerable 68\% judged the consideration of all main disciplines of physics to be very important which emphasises the relevance of a spectral eERP. Figure \ref{fig:euroq} shows four questions concerning the eEERP concept to investigate the Eurogame concept by VP, reflecting the success of the eEER concept, for details see below.
\section{Discussion}
\subsection{Questions and Tests}
\begin{figure}
    \centering
    \includegraphics[width=0.9\textwidth]{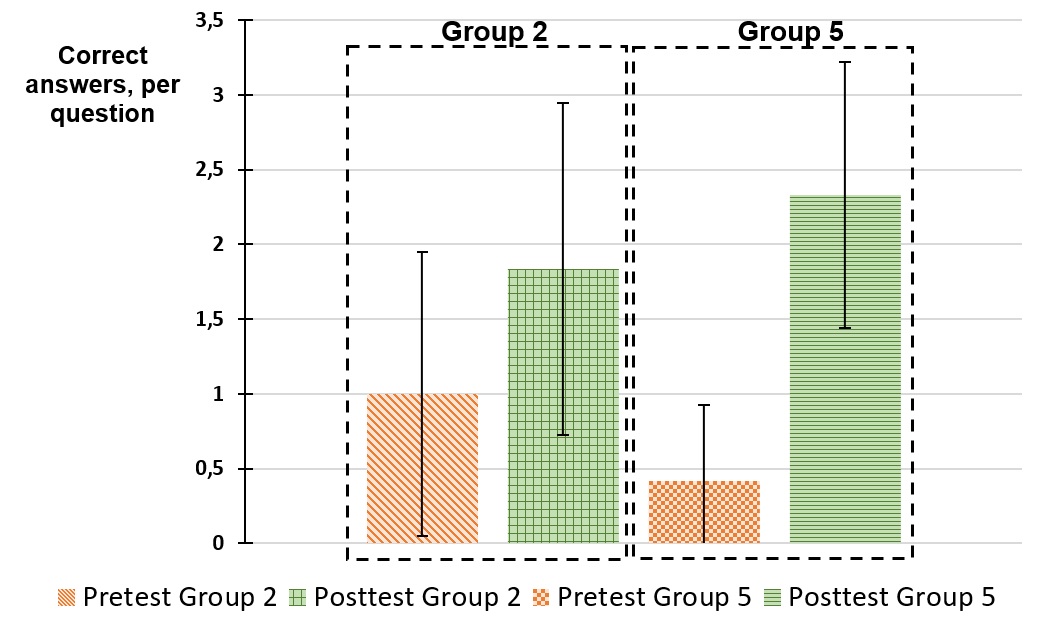}
    \caption{Statistical evaluation based on the total number of questions ($N_3=12$), group-separated and averaged numbers of correct answers per question of pre- (orange) and posttest (green) in comparison.}
    \label{fig:groups}
\end{figure}

In general, an effort was made to find questions which could not be answered purely from the lecture so as to be able to measure the increase in knowledge, e.g., 3 students out of 25 could answer the 60-Hz-question above, avoided too high a zero-rate and wasting data points for statistics. Therefore, the questions were very specific which was retrospectively a successful strategy. We asked two questions per discipline in physics. Exceptionally, thermodynamics (entropy, heat engine, fridge, efficiency) covered four questions associated with one knowledge transfer card and PP. These four topics are less discussed in the physics lecture being a measure of knowledge which clearly exceeded the lecture. The analysis also indicated the dependence of this success on the individual student group due to group inhomogeneities in willingness to learn, learning ability and group dynamics. Exemplarily, Figure \ref{fig:groups} shows a comparison of two groups of 3 players (group separated treatment; for details, see \textit{Supplementary}). Group 5 showed a higher level of significance of knowledge increase. The increase is significant for the question of entropy, heat engine and fridge. The question concerning efficiency of a heat engine had a low value of correctness. It is worth mentioning that, the puzzle Stirling engine was the last PP in the eEERP. Assuming a decrease in concentration among the players, the success of the increase was counted to be positive. Probably the notation of the last answers $\eta<1$, $\eta=1$ or $\eta>1$ was difficult to cope with and to remember after the first confrontation without a deep understanding. Statistically, the last question can be counted as an outlier. The eEER-test contained two kinds of questions: general eER questions and EER questions. A considerable 80\% rated a room concept to be very important, probably seen as a break, some sort of adventure. The EPs and the VP system were rated as good (80\%) and it could motivate players (87\%), significantly felt challenged (76\%). Therefore, the concept of VP was considered successful.
\subsection{Educational Escape Rooms for Physics}
C. Lathwesen \textit{et al.} found just six protocols of eERP \cite{educsci11060308}. All targeted secondary education focusing on a specific topic. According to C. Lathwesen \textit{et al.}: 
\begin{quotation}\textit{Nevertheless, we were able to identify some gaps in the research and development on this topic. First of all, there is still a need for new proposals mainly in the subjects of
physics and biology} \cite{educsci11060308}.
\end{quotation}
 Our contribution follows this request. Interestingly, spectral eERP are extrem discussed. Most of them covered a narrow topic/discipline. Additionally, l. Hayden \textit{et al.} requested:
 \begin{quotation}\textit{Overall, the reviewed articles indicate that more studies are needed on the use of escape rooms for educational purposes. It seems that research on this topic has reached a new phase, and there is a need for structured research and transparency in the research design and methods for data collection and analysis} \cite{doi:10.1080/20004508.2020.1860284}.
\end{quotation}
It is unconventional to discuss the topic of transparency and structure in a paper here because it should be self-evident. However, the above citation shows it is not. In physics and natural science, research is very structured and transparent. Therefore, the effort was made to choose a structure as is usual in physics/chemistry publications to cover the range of topics (Eurogame strategies and broad range of disciplines in physics). Probably the above criticism is a result of the interdisciplinary field involving various subjects and writing styles, preparing an eER just secondarily in parallel with other research. In this work, various ways of data analysis have been presented for scrupulous interpretation.
\subsection{Comparison of eEERP and Eurogames}
The common advantages of eERs prevail. The mandatory sequence of PPs requires prioritisation. This is an improvement, adding decision-making and setting priorities for the list of promoted skills of eERs. This is an advantage of the eEERP together with the fun-factor of the EPs. The adapted main point is the VP system of Eurogames which allows to introduce puzzles without a strong topical connection to the subject of interest, in this case physics. EPs have a character of pure entertainment, exceptions such as argumentation that dexterity cubes are able to train surgical dexterity being excluded. The player interaction can be understood to be like a cooperative Eurogame. Groups of players of various teams do not interact, just for clarity. We are strongly propose adopting the ideas of EERs including EPs and applying the here presented concept to other eER. The score list of VP is able to be reset, the decision for every new run being left to the e-game master.

\section{Conclusions}
We presented an eEERP instrumentalised Eurogame concepts of VP as a method of motivation for knowledge transfer in physics. We combined a strategy of repetition and application of knowledge gained from the lecture (main part, 3/4) and the introduction of new knowledge (1/4), mainly in the field of thermodynamics and topics strongly related with the main field of the players, in this case veterinary medicine, e.g., the name of proteins associated with the orientation of birds in the magnetic field of earth. An increase in knowledge was verified together with positive feedback on the Eurogame concept. Players have been motivated and challenged as shown by the test groups. By the eEER, the list of advantages promoting skills can be extended to setting priorities because  of a conflict of choice between adventure-like EPs and PPs. However, PPs are mandatory to master the eEERP. An advantage is the flexibility of the European board concept of VP, choosing from of a range of optional EPs to fascinate and captivate students and at the same time taking into account individual preferences. The Eurogame concept using arbitrary EPs to collect VP can be applied to every eER and our findings support its implementation. \faDice

\section*{Acknowledgments}
Special thanks go to Elisabeth Schaper and Christin Kleinsorgen (University of Veterinary Medicine Hannover Foundation)  for helpful discussions regarding the project proposal. The project was supported by the program \textit{InnovationPlus} (Niedersächsisches Ministerium für Wissenschaft und Kultur).

\section*{Authors declaration}
The authors have no conflicts to disclose.
\clearpage
\section*{References}
\bibliographystyle{unsrt.bst}
\bibliography{bib}

\appendix

\section{List of victory points}
Table \ref{jlab4} shows the victory points $P_i$ of PP and points $p_i$ of EPs as used in our EERP in 2023/2024. Statistically, the students earned $\approx20.1$ VP per player visited the eEERP. 
\begin{table}
\caption{\label{jlab4}Overview of VP of the PPs and EPs.}
\centering
\footnotesize
\begin{tabular}{@{}lcc}
\br
Puzzle&Victory points of PPs: $P_i$&Victory points of EPs: $p_i$\\
\mr
Nobel prize (start-up)&2&\\
Apple puzzle&6&\\
Fluorescence&3&\\
Electric conductivity&4&\\
Total internal reflection&3&\\
50 Hertz sound experiment&4&\\
X-ray&4&\\
Dog puzzle&8&\\
Train experiment \cite{tiho_mods_00010208}&4&\\
Cup Stirling machine&2&\\
\mr
Backgammon&&3\\
Biochemistry&&3\\
Domino&&2\\
Einstein lock&&4\\
Bottle safe&&5\\
24 dexterity cubes&&24 total, 1 per cube\\
6 Wood puzzle&&24, 4 per puzzle\\
Card game&&4\\
10 knots&&10, 1 per knot\\
Magnet blocks&&6\\
Periodic table&&3\\
Wood bottle&&4\\
Chess&&7\\
3x3 square&&5\\
Text challenges&&20 total, 4 per text card\\
Hidden card&&2\\
\br
\end{tabular}\
\end{table}
\normalsize
\newpage
\section{Supplements of the magnetic dog}
\begin{figure}
    \centering
    \includegraphics[width=\textwidth]{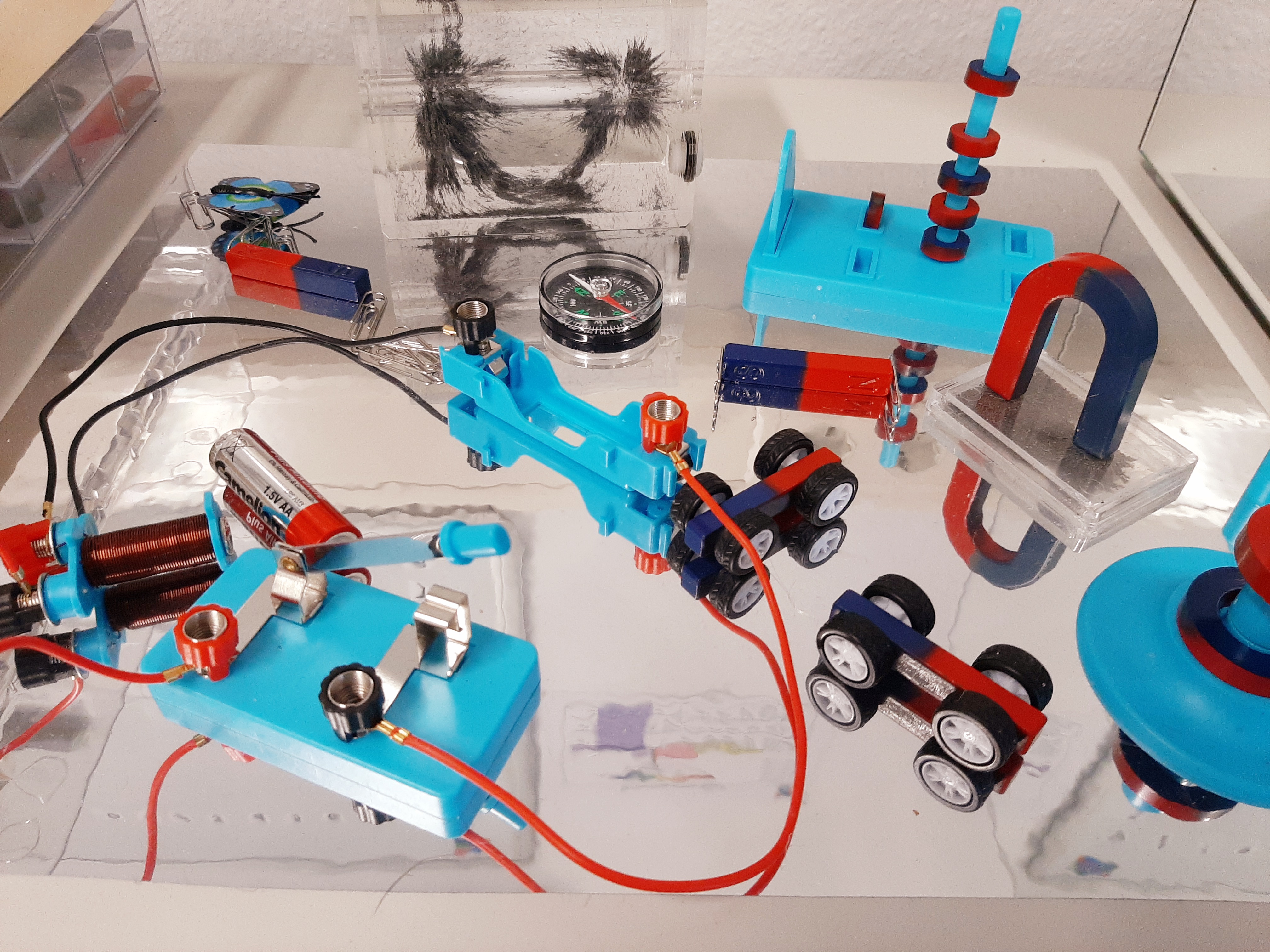}
    \caption{Small experiments with magnets located on mirror foil to support the dog puzzle. Keep in mind that, the transparent box in the background contains an oil-Fe-particle suspension to reveal magnetic field lines. It is not part of the presented set, belonging to of the equipment used at physics of the university.}
    \label{fig:magnetstuff}
\end{figure}

We supplemented small magnetic experiments, being an add-on without any VP. This is shown in Figure \ref{fig:magnetstuff} presenting a commercial solution \textit{Magnets Experiment Set} from \textit{Burnur Science}. Various conventional ferromagnets are colourised to visualize North- and South Pole. The set contains
\begin{itemize}
    \item[\faMagnet] two bar magnets with wheels: magnet cars,
    \item[\faMagnet] an electromagnet with paper clips to test it,
    \item[\faMagnet] two sets of ring magnet levitated by forces,
    \item[\faMagnet] a compass and a horseshoe magnet,
    \item[\faMagnet] iron fragments, magnetic butterflies and a manual.
\end{itemize}
\end{document}